\documentclass[aps,preprint,nofootinbib,floatfix,a4paper]{revtex4-1}
\pdfoutput=1
\usepackage{amsmath}
\usepackage{graphicx}
\usepackage{subfigure}
\usepackage{epstopdf}
\usepackage{epsfig}
\usepackage{amssymb}
\usepackage{bm}
\usepackage{bbm}
\usepackage{slashed}
\usepackage{hyperref}
\newcommand{\beq}{\begin{equation}}
\newcommand{\eeq}{\end{equation}}

\usepackage{color}

\begin{document}

\title{Probing a new strongly interacting sector\\ 
via composite diboson resonances}

\author{P. Ko}
\email[]{pko@kias.re.kr}
\affiliation{School of Physics, KIAS, Seoul 130-722, Korea}

\author{Chaehyun Yu}
\email[]{chyu@korea.ac.kr}
\affiliation{Institute of Physics, Academia Sinica, Nangang, Taipei 11529, Taiwan}
\affiliation{ Department of Physics, Korea University, Seoul 02841, Korea}

\author{Tzu-Chiang Yuan}
\email[]{tcyuan@gate.sinica.edu.tw}
\affiliation{Institute of Physics, Academia Sinica, Nangang, Taipei 11529,
Taiwan}

\date{\today}

\begin{abstract}
\noindent
Diphoton resonance was a crucial discovery mode for the 125 GeV Standard Model Higgs boson
at the Large Hadron Collider (LHC).
This mode or the more general diboson modes may also play an important role 
in probing for new physics 
beyond the Standard Model.
In this paper, we consider the possibility that a diphoton resonance 
is due to a composite scalar or pseudoscalar boson, 
whose constituents are either new hyperquarks $Q$ or scalar hyperquarks 
$\widetilde{Q}$ confined by a new hypercolor force at a confinement scale $\Lambda_h$. 
Assuming the mass $m_Q$ (or $m_{\widetilde Q}$) $\gg \Lambda_h$, 
a diphoton resonance could be interpreted as
either a $Q\overline{Q} (^1S_0)$ state $\eta_Q$ with $J^{PC} = 0^{-+}$ or a
$\widetilde{Q} \widetilde{Q}^\dagger (^1S_0)$ state $\eta_{\widetilde Q}$ 
with $J^{PC}=0^{++}$. 
For the $Q\overline{Q}$ scenario, there will be a spin-triplet partner $\psi_Q$ which is slightly 
heavier than $\eta_Q$ due to the hyperfine interactions mediated by hypercolor gluon exchange;
while for the $\widetilde{Q} \widetilde{Q}^\dagger$ scenario, the spin-triplet partner $\chi_{\widetilde Q}$  
arises from higher radial excitation with nonzero orbital angular momentum.
We consider productions and decays of $\eta_Q$, $\eta_{\widetilde Q}$, 
$\psi_Q$, and $\chi_{\widetilde Q}$
at the LHC using the nonrelativistic QCD factorization approach.
We discuss how to test these scenarios by using the Drell-Yan process
and the forward dijet azimuthal angular distributions to determine the 
$J^{PC}$ quantum number of the diphoton resonance. 
Constraints on the parameter space can be obtained by interpreting some of the small diphoton ``excesses'' reported by the LHC as the composite scalar or pseudoscalar of the model.
Another important test of the model 
is the presence of a nearby hypercolor-singlet but color-octet state like the $^1S_0$ state $\eta^8_Q$
or $\eta^8_{\widetilde Q}$, which can also be constrained by dijet or monojet plus monophoton data.
Both possibilities of a large or small width of the resonance can be accommodated, depending on whether 
the hyper-glueball states are kinematically allowed in the final state or not.
\end{abstract}

\maketitle

\section{Introduction}
\label{sec1}

It is well known that the discovery mode of the 125 GeV Standard Model (SM) Higgs boson 
at the Large Hadron Collider (LHC) is the diphoton channel, $h_{\rm SM}(125 \, {\rm GeV}) \to \gamma\gamma$. Perhaps it is somewhat ironic that the discovery mode of the Higgs boson has something to do with the one-loop induced higher-dimension operator\footnote{The dominant production mechanism for the SM Higgs at the LHC 
is also a one-loop induced gluon fusion process.}
for the diphoton mode, rather than the other renormalizable tree-level vertices coming from the spontaneous breaking of the electroweak gauge symmetry via the Higgs mechanism.
All such tree-level couplings, $hf\bar f$ and $h VV$ ($f$ and $V$ denote the SM fermion and weak gauge boson, respectively), are proportional to the masses of the final-state particles, which is a generic feature of the Higgs mechanism. Thus, for a relatively light 
SM Higgs of 125 GeV, all kinematically accessible tree-level processes are 
suppressed by the light masses of the final-state particles. It is therefore 
of the upmost importance for LHC run II (as well as for future $e^+e^-$ colliders like CEPC or ILC) 
to verify that the 125 GeV boson does couple to the SM fermions and that weak gauge bosons are in line with SM expectations. 

Nevertheless, the diphoton mode remains an important process, since  it is a very clean signal at the LHC. 
In particular,  this mode may play a role for probing new physics beyond the SM. Recall that the 750~GeV bump 
reported around Christmas time in 2015 by the two LHC collaborations~\cite{Aaboud:2016tru,Khachatryan:2016hje} 
is also a diphoton resonance. This bump was very hard to explain within SM, 
and many different ideas have been proposed to accommodate this.
Unfortunately, it was rather short lived -- the ``excess'' has faded away in the summer
after more data were collected and analyzed~\cite{ATLAS:2016eeo,Khachatryan:2016yec}.

It is not necessary for diphoton resonance to arise from one-loop induced  amplitude.
An attractive alternative scenario is to introduce a composite bound state of new heavy particles
with QCD and/or QCD-like interactions, as was considered in Refs.~\cite{
Luo:2015yio,Cline:2015msi,Craig:2015lra,Han:2016pab,Kats:2016kuz,Kamenik:2016izk,
Iwamoto:2016ral,Anchordoqui:2016ogi,Foot:2016llc}
to explain the ``excess'' of the 750~GeV bump.
Here the diphoton amplitude is not suppressed by the loop but rather
by the wave function for finding two heavy particles at the origin 
to form the bound state.
This scenario is distinguishable from another interesting scenario, where
the diphoton excess is due to 
a pseudo--Nambu-Goldstone boson (pNGB) coming from the spontaneous
breakdown of a global symmetry~\cite{Harigaya:2015ezk,Nakai:2015ptz,%
Matsuzaki:2015che,Belyaev:2015hgo,Harigaya:2016pnu,Harigaya:2016eol,%
Chiang:2015tqz,Barrie:2016ntq,Nevzorov:2016fxp,Cline:2016nab,Bai:2016czm},
and the diphoton amplitude is suppressed by the anomaly term.
In general, the new composite states can be investigated through
any diboson resonance as well as diphoton resonance 
at the LHC~\cite{Arbey:2015exa,Cacciapaglia:2015eqa,Cacciapaglia:2015nga,%
Carpenter:2015gua,Cai:2015bss,Ferretti:2016upr,Belyaev:2016ftv,Englert:2016ktc}.

In this paper, we explore in detail such a scenario in which diphoton (or in general, diboson) resonance 
that might appear in the future LHC experiments 
may be due to new confining strong interaction
(which we call hypercolor interaction, or h-QCD in short)
and new particles (h-quark $Q$ or scalar h-quark $\widetilde Q$) 
that feel not only this new strong force but also the SM gauge interactions.
If the new particles belong to a $SU(2)_L$ doublet and feel strong color interactions, it would modify
the 125~GeV Higgs signal strength in the $gg\rightarrow H \rightarrow \gamma\gamma$ channel.
And there would be strong constraints from electroweak precision tests parametrized by the oblique $S$ and $T$
parameters. To avoid these issues, we assume that the new particles are colored but $SU(2)_L$ singlets with hypercharge $Y=e_Q$.\footnote{ In the numerical analysis, we will take $Y=e_Q = 2/3$, 
and  one can easily scale the results for other values of $Y=e_Q$.}
We consider the spin of the new particle being either 0 (complex scalar boson $\widetilde Q$) or 1/2 (Dirac fermion $Q$)
and study their lowest-lying bound states, $\eta_{\widetilde Q}(^1S_0)$, 
$\eta_Q(^1S_0)$, and $\psi_Q(^3S_1)$.

For the case where the new fermion $Q$
belongs to a $SU(2)_L$ doublet but feels no strong color interaction, as was 
discussed previously in the context of quirks~\cite{Kang:2008ea} or iquarks~\cite{Cheung:2008ke}, 
besides the $\gamma\gamma$, $ZZ$, and $Z\gamma$ channels, 
other diboson decay modes of the hyperquarkonia 
like $W^+W^-$, $W^\pm\gamma$, and $W^\pm Z$ in the final states 
are also possible.
A more general case for the heavy fermion $Q$ being a colored $SU(2)_L$ doublet will be treated in Ref.~\cite{progress}.

The paper is structured as follows:~In Sec.~\ref{sec2}, we set up the model of hypercolor QCD and discuss its bound-state spectra, 
including the $^1 S_0$ color-octet states $\eta^8_Q$ and $\eta^8_{\widetilde Q}$.
The productions and decays of the bound states at the LHC 
for the vectorlike hyperquark and the scalar hyperquark cases 
are discussed in Secs.~\ref{sec3} and \ref{sec4}, respectively.
In Sec.~\ref{sec5}, 
we briefly discuss how to distinguish between the two scenarios of hyperquark and hyperscalar quark composites.
In Sec.~\ref{sec6}, we discuss the possible interpretation of the high-mass diphoton resonances at 710~GeV and 1.6~TeV reported with small ``excesses'' at the LHC as a composite scalar $\eta_{\widetilde Q}$ or pseudoscalar $\eta_Q$ in the model.
We also briefly discuss the small ``excess'' of the photon + jet resonance at 2~TeV as the decay product of the color-octet state $\eta^8_Q$ or $\eta^8_{\widetilde Q}$. 
Finally, we summarize our study in Sec.~\ref{sec7}.

\section{Hypercolor Model Setup}
\label{sec2}

For the hyper--strongly interacting model, 
we assume that 
\begin{itemize}
\item[(1)]
There is a new confining gauge group $SU(N_h)$ with strong coupling $g'$ 
and a confinement scale $\Lambda_h$, defined as 
\begin{equation}
\Lambda_h \simeq M {\rm exp} \left[  - \frac{6\pi}{(11 N_h - 2 n_{f^h}) \alpha_h (M)}
\right] \, ,
\end{equation}
where $n_{f^h}$ is the number of hyperquark flavors, $M$ is a heavy mass scale, and
$\alpha_h = g'^2/4 \pi$.
\item[(2)]
There is a new vectorlike h-quark (hyperquark) $Q$ and its antiparticle $\overline{Q}$
(or scalar h-quark $\widetilde{Q}$ and its antiparticle $\widetilde{Q}^\dagger$),  
whose quantum numbers under the $SU(3)_C \times SU(2)_L \times U(1)_Y \times SU(N_h)$ 
are defined as $( 3, 1, Y ; N_h )$.
\item[(3)]
Both $Q$ and $\widetilde{Q}$ are heavier than the 
confinement scale $\Lambda_h$, so that $Q\overline{Q}$ 
($\widetilde{Q}\widetilde{Q}^\dagger$) bound states can be treated as 
heavy hyperquarkonia, analogous to $J/\psi, \eta_c$, $\Upsilon, \eta_b$, etc.~in QCD. 
\end{itemize}

If $\alpha_h ( m_Q v_Q ) m_Q > \Lambda_h$, the bound system would be more like a 
Coulombic bound state, since the nonperturbative confinement effect would be smaller 
than the Coulomb interaction.   
One can show that Coulomb dominance can be a reasonably good approximation
for the entire range of $\alpha_h$~\cite{progress}.
In the following, we will accept this assumption  and present various numerical results assuming the binding potential $V$ is Coulombic. Namely,
\begin{equation}
V = - \frac{C_h \alpha_h}{r} -\frac{C_F \alpha_s}{r},
\end{equation}
with $C_h = ( N_h^2 - 1)/(2 N_h)$ and $C_F=(N_c^2-1)/(2 N_c)$. 
Note that the new strong interaction dominates over QCD interaction 
for $\alpha_h(M) \gtrsim 0.2$, while the two interactions are competitive
with each other for $\alpha_h(M) \sim 0.1$.
When interpreting the results, one has to keep in mind 
that these numerical results are based on the assumption of Coulomb dominance.  
The wave function at the origin for the radial quantum number $n=1$ $S$-wave ground state
assuming Coulomb dominance is given by~\cite{coulomb}
\begin{equation}
| R_{1S} (0) |^2 = m_Q \biggl\langle \frac{dV}{dr} \biggr\rangle = 
4 \left( [C_h \alpha_h + C_F \alpha_s]\frac{m_Q}{2}
\right)^3 .
\label{wavefunction}%
\end{equation}
This nonperturbative  quantity is very important, since it determines both production and decay 
rates of the $S$-wave $Q\overline{Q}$ bound states.
The wave function $\widetilde R_{1S}(0)$ for the $\widetilde Q \widetilde Q^\dagger$ bound state
is approximately the same as $R_{1S} (0)$, up to the one-loop correction to the hyper-QCD potential~\cite{Moxhay:1985dy}.

Besides the heavy $Q$, there is also the massless h-gluon $g_h$.
Due to h-color confinement, the lightest h-hadron would be a 
scalar or pseudoscalar h-glueball state.
For pure $SU(3)_h$ case, the lightest scalar glueball mass is given by 
$m_0 \sim (4\textrm{-}7) \Lambda_h$~\cite{Chen:2005mg,Gregory:2012hu,Juknevich:2009gg}.
Depending on the mass of the h-glueball, the lightest $Q\overline{Q}$ (or 
$\widetilde{Q} \widetilde{Q}^\dagger$) bound state 
may or may not decay into two h-glueballs.
In this work, we consider cases where decay into h-glueballs is either
open or forbidden kinematically.

\subsection{Spectra of new resonances}

We assume that $\alpha_h ( m_Q v_Q ) \sim v_Q^2 \ll 1$, so that the h-QCD version of 
nonrelativistic QCD (NRQCD)~\cite{nrqcd} for charmonia and bottomonia applies.  Otherwise, there is no systematic way to calculate
decay and production rates for the bound states.  This condition implies that 
if $\alpha_h (M) \sim 0.5$ or larger, then the system would no longer be nonrelativistic, and
there is no guarantee that the NRQCD approach would give a good description of $Q\overline{Q}$ bound
states.   
As mentioned before, we also assume $\alpha_h M \gg \Lambda_h$, so that the nonperturbative effects 
are small and one can make an approximation using the Coulomb potential for the 
$Q\overline{Q}$  system.  Then the binding energy of this system is approximately given by
\begin{equation}
M ( n^{2S+1}L_J ) \simeq 2 m_Q  \left[ 1 - \frac{(C_h \alpha_h+C_F \alpha_s)^2}{8 n^2} \right].
\end{equation}
Note that the degeneracy in the orbital quantum number $l$ is special only for the 
Coulomb potential.   
The mass of the lowest state, $\eta_Q$, is approximately given by
$M_{\eta_Q} = M(1 ^1S_0 ) \approx 2 m_Q$ for small $\alpha_h$.
The excited $2^1S_0$ state $\eta^{\prime}_Q$ has a mass
\begin{equation}
M ( \eta^{\prime}_Q ) = 2 m_Q\left( \frac{1- [C_h \alpha_h +C_F\alpha_s]^2 / 32}{
1- [C_h \alpha_h +C_F\alpha_s]^2 / 8} \right) .
\end{equation}
For instance, for $\alpha_h = 0.2$ and $m_Q=1$~TeV, the mass difference of $\eta^\prime_Q$ and $\eta_Q$
is about $28$, $47$, and $70$ GeV for $N_h=3$, $4$, and $5$, respectively.

The mass of a spin-triplet partner $\psi_Q$ is determined by hyperfine splitting
\begin{equation}
\frac{M_{\psi_Q} - M_{\eta_Q}}{M_{\eta_Q}} = \frac{16 \pi}{3} \alpha_h \frac{| R_S (0) |^2}{M^3}
\approx \frac{\pi}{3 n^2} ( C_h \alpha_h+C_F\alpha_s)^4 ,
\end{equation}
where the last equation only holds for Coulomb potential between $Q$ and $\overline{Q}$.
The resulting mass splitting between $^1S_0$ and $^3S_1$ is
\begin{equation}
\Delta M \lesssim (45, 122, 264) ~{\rm GeV \; for}~N_h = (3,4,5).
\end{equation}
For simplicity, we ignore the mass difference and set $M_{\psi_Q} = M_{\eta_Q}$
in our analysis. 

In the scalar h-quark scenario, we expect that the mass spectrum
of low-lying states are the same as that in the h-quark case up to one-loop
correction and spin-dependent hyperfine splitting,\footnote{ 
The hyperfine splitting is proportional to 
$1/m_Q^2$, so that it would be negligible for heavy h-quarks.} because the potentials in the two scenarios 
are identical.

\subsection{Color-octet bound state}

Next, we consider the  $Q\bar{Q} (^1S_0)$ bound state, $\eta_Q^8$, which is a singlet
under h-QCD, but an octet under ordinary QCD.
One can easily extend the analysis to other color-octet states with different spin and orbital angular momentum.
It is well known that the potential of a $Q\bar{Q}$ pair is attractive
in the color-singlet state, but repulsive in the color-octet state.
Nevertheless, the $\eta_Q^8$ bound state can still be formed because the attractive hyper--strong  
interaction is stronger than the repulsive one from ordinary QCD.
The potential of the $Q\bar{Q}$ pair is expressed as the sum of two terms
\begin{equation}
V = - \frac{C_h \alpha_h}{r} + \frac{C_8 \alpha_s}{r},
\end{equation}
where $C_8= C_A/2 - C_F$ with $C_A=N_c$. The wave function $R_{\eta_Q}^8 (0)$ at the origin
of $\eta_Q^8$ can be given in the same form as Eq.~(\ref{wavefunction})
by the substitution of $C_h \alpha_h +C_F\alpha_s\rightarrow C_h \alpha_h - C_8 \alpha_s$.

Similarly, one can obtain the wave function at the origin,
${\widetilde R}^8_{\eta_{\widetilde Q}}(0)$, for the scalar h-quark pair.

\section{Bound states of hyperquarks}
\label{sec3}

In this section, we consider a vectorlike h-quark singlet $Q $ with $Y=e_Q=2/3$ 
and mass $m_Q$.  $Q$ belongs to the fundamental representations of both $SU(N_h)$ and 
ordinary $SU(3)_C$ gauge theories, and thus feels new strong interaction as well as ordinary strong 
interaction.     First, we consider the spin-singlet $S$-wave state $\eta_Q(^1S_0)$.
Then, the spin-triplet $S$-wave state $\psi_Q(^3S_1)$ will be taken into account.

\subsection{Production and decay of $\eta_Q$}

The pseudoscalar bound state $\eta_Q$ of new hidden quarks can decay into
two photons, $\gamma Z$, $ZZ$, two gluons, or two h-gluons.
Their decay widths are given by
\begin{eqnarray}
\Gamma (\eta_Q \to \gamma\gamma)&=&
\frac{N_c N_h \alpha^2 e_Q^4}{m_Q^2}
 \left|R_{1S}(0)\right|^2,
\\
\Gamma (\eta_Q \to gg) &=&
\frac{C_F N_h \alpha_s^2}{2 m_Q^2}
\left|R_{1S}(0)\right|^2,
\\
\Gamma(\eta_Q\to \gamma Z) & = & (x_w (4-r_Z)/2 (1-x_w)) \Gamma(\eta_Q\to \gamma\gamma), 
\\
\Gamma(\eta_Q \to ZZ) &= & 
\frac{4 N_c N_h \alpha^2 e_Q^4 x_w^2 (1-r_Z)^{3/2}}{m_Q^2 (2-r_Z)^2 (1-x_w)^2}
 \left|R_{1S}(0)\right|^2,
\\
\Gamma(\eta_Q \to g_h g_h) & = & (C_h N_c \alpha_h^2 / C_F N_h \alpha_s^2) \Gamma(\eta_Q \to gg).
\end{eqnarray}
Here $x_w = \sin^2 \theta_W$ and $r_Z = m_Z^2/4m_Q^2$.
We note that $\eta_Q$ does not decay into a pair of fermions or $WW$ owing to
the singlet nature of $Q$ and the $J^{PC}$ quantum number of $\eta_Q$ being  $0^{-+}$.\footnote{
We shall ignore loop-induced decays such as 
$\eta_Q \rightarrow  \gamma^* \gamma^* , Z^* Z^* , Z^* \gamma^*  \rightarrow f\bar{f}, W^+ W^-$, 
because they are loop suppressed.}
The branching ratios strongly depend on $\alpha_h$ if $\eta_Q\to g_h g_h$ is
allowed. For $\alpha_h \sim 0.1$, $BR(\eta_Q\to g_h g_h)\sim BR(\eta_Q\to gg)\sim 0.5$. 
However, for $\alpha_h \gtrsim 0.2$, the $\eta_Q \to g_h g_h$ channel is
dominant. 
If $\eta_Q\to g_h g_h$ is kinematically forbidden, $BR(\eta_Q \to gg)$ becomes
$0.99$ irrespective of $\alpha_h$ and $N_h$~\cite{progress}.

At the LHC, the $\eta_Q$ can be produced via gluon fusion.
The cross section for the diphoton production $pp\to \eta_Q \to \gamma\gamma$
is given by
\begin{equation}
\sigma(pp \to \eta_Q \to \gamma\gamma) =
\frac{C_{gg}}{s M_{\eta_Q} \Gamma_\textrm{tot}}
\Gamma ( \eta_Q \to gg )~ \Gamma (\eta_Q \to \gamma\gamma ),
\label{ggxsec}%
\end{equation}
where $C_{gg}$ is defined as~\cite{Franceschini:2015kwy}
\begin{equation}
C_{gg}=\frac{\pi^2}{8}\int_{M^2/s}^{1} \frac{d \tau}{\tau} 
f_g(\tau)f_g(M^2/s\tau)
\end{equation}
with $f_g(\tau)$ being the gluonic parton distribution function  
at the longitudinal momentum fraction of the gluon $\tau$.
By making use of the MSTW2008NLO data at $\sqrt{s}=13$ TeV~\cite{mstw}, 
one finds that $C_{gg}=2137$ and 7.14 at $M=750$ and 2 TeV,
respectively.
Similarly, one can obtain the cross section for the two-gluon production via
the $\eta_Q$ decays.

\begin{figure}[t]
\includegraphics[width=0.45\textwidth]{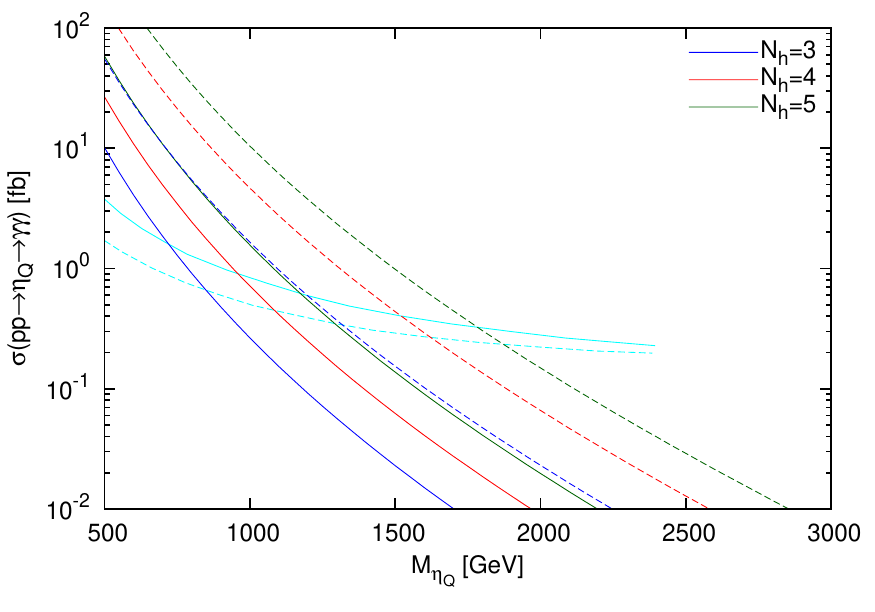}
\includegraphics[width=0.45\textwidth]{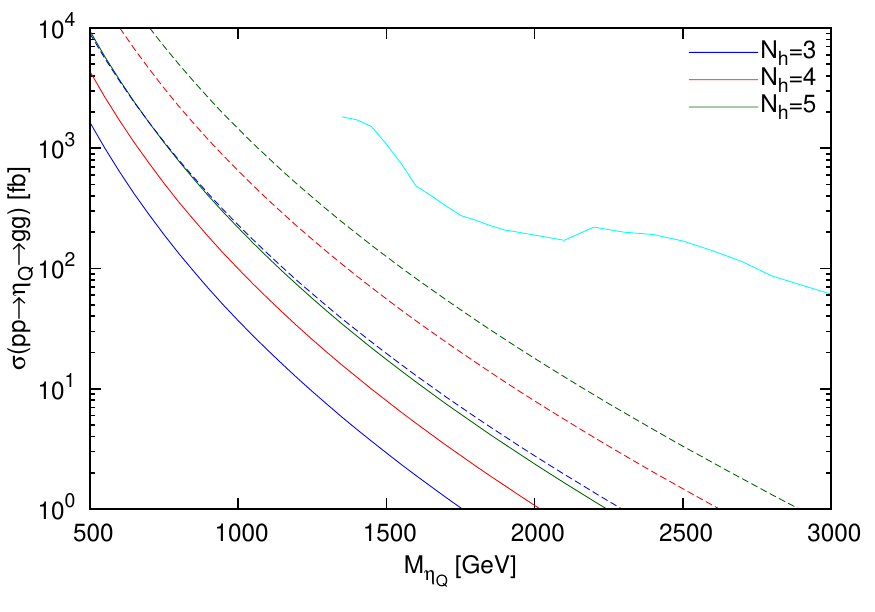}
\includegraphics[width=0.45\textwidth]{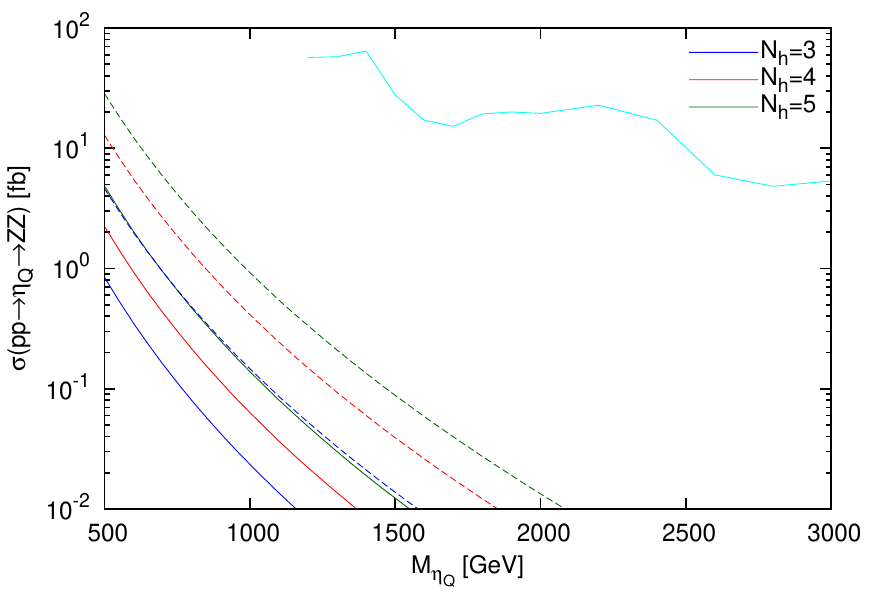}
\includegraphics[width=0.45\textwidth]{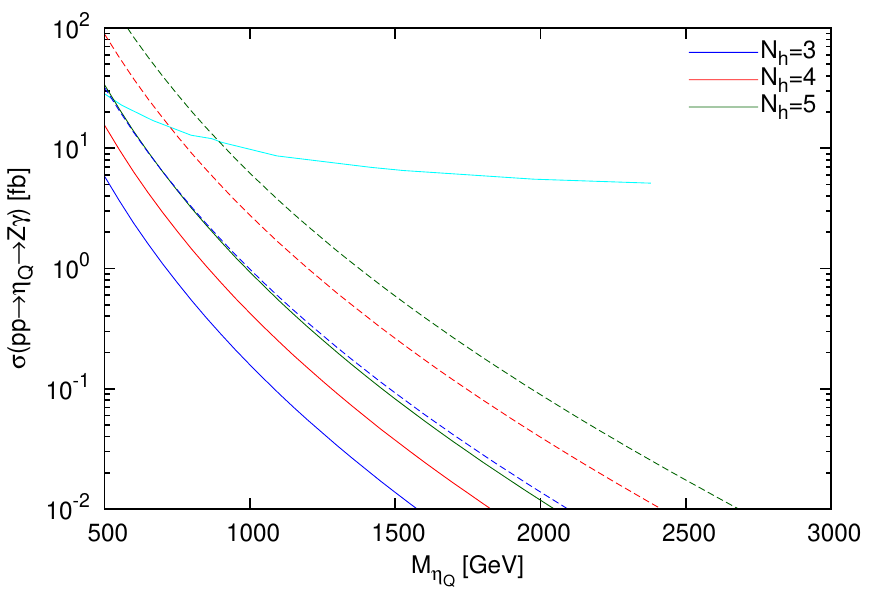}
\caption{
The cross sections for (a) $pp\to \eta_Q \to \gamma\gamma$,
(b) $pp\to \eta_Q \to gg$, (c) $pp\to \eta_Q \to ZZ$,
and (d) $pp\to \eta_Q \to Z\gamma$ for $\alpha_h=0.2$
at the LHC with $\sqrt{s}=13$ TeV in units of fb as functions of $M_{\eta_Q}$.
The blue, red, and green solid (dashed) lines 
in these plots correspond to the $N_h=3$, $4$, and $5$ cases, respectively,
in which $\eta_Q\to g_h g_h$ is allowed (forbidden).
See the text for explanation of the cyan lines in each of these plots.
}
\label{figmq}
\end{figure}

In Fig.~\ref{figmq}, we show the production cross sections
of (a) two photons, (b) two gluons, (c) two $Z$ bosons, and (d) $Z\gamma$
via the $\eta_Q$ resonance 
for $\alpha_h = 0.2$ as functions of the mass of $\eta_Q$, $M_{\eta_Q}$.
The blue, red, and green lines denote the $N_h=3$, $4$, and $5$ cases, respectively,
where the solid (dashed) lines correspond to the cases in which 
the $\eta_Q\to g_h g_h$ channel is open (closed).

In Fig.~\ref{figmq}(a), the solid (dashed) cyan line represents the expected 
95\% C.L. upper limit on the fiducial cross section times the branching ratio 
of a spin-0 resonance to two photons
at $\sqrt{s}=13$~TeV in ATLAS data by assuming the ratio $\Gamma/M_{\eta_Q}=2$ \%
($\Gamma=4$~MeV)~\cite{ATLAS:2016eeo}.
We note that the observed 95\% C.L. upper limit in ATLAS is 
almost the same as the one expected by the ATLAS Collaboration~\cite{ATLAS:2016eeo}. 
Since the total decay width of $\eta_Q$ is about $150$~MeV to $10$~GeV
for $\alpha_h=0.2$, one could impose the bound on the model somewhere
between the two cyan lines. 
Note that the ratio $\Gamma/M_{\eta_Q}$ could be about 10 \% for larger $\alpha_h$.
As shown in Fig.~\ref{figmq}(a), the lower bound on $M_{\eta_Q}$ is 
about $800$ ($1200$)~GeV for $N_h = 3$ ($5$) 
if $\eta_Q\to g_h g_h$ is allowed, while 
it could be about $1300$ ($1900$)~GeV if $\eta_Q\to g_h g_h$ is closed.
The difference for the lower bounds simply arises from
the difference in the total decay width of $\eta_Q$, which is much larger
in the former case.

In Figs.~\ref{figmq}(b)--\ref{figmq}(d), we show the cross sections for
(b) $pp\to \eta_Q\to gg$, (c) $pp\to \eta_Q\to ZZ$,
and (d) $pp\to \eta_Q\to Z\gamma$.
The cyan lines denote the observed
95\% C.L. upper limits on the fiducial cross section times branching ratio 
for (b) dijet production~\cite{atlasjj}, (c) $ZZ$ production~\cite{atlaszz}, 
and (d) $Z\gamma$ production~\cite{atlaszp} at $\sqrt{s}=13$ TeV in ATLAS data.
As shown in Fig.~\ref{figmq}, the $gg$ and $ZZ$ productions are not constrained
by experiments yet. However, the search for a resonance which
decays into $Z \gamma$ starts by constraining this model,
in particular, in the case that $\eta_Q\to g_h g_h$ is forbidden. 

In summary, the case of $pp\to \eta_Q\to \gamma\gamma$ is mostly constrained
by current experimental data. In other words, $\eta_Q\to \gamma\gamma$
would be the most promising channel for probing this composite model. 
One may obtain similar results with experimental bounds at $\sqrt{s}=8$ or $13$~TeV 
in CMS or ATLAS for $pp\to \gamma\gamma$~\cite{2gamma1,2gamma2,2gamma3},
$pp\to jj$~\cite{gg1,gg2,gg3,gg4,gg5},
$pp\to ZZ$~\cite{zz1,zz2,zz3,zz4,zz5,zz6,zz7,zz8,zz9,zz10,zz11}, and
$pp\to Z\gamma$~\cite{zgamma1,zgamma2,zgamma3,zgamma4,zgamma5}.

\subsection{Production and decay of $\psi_Q$}

One of the decisive tests for a spin-singlet 
$S$-wave bound state $\eta_Q$ of a new fermion-antifermion pair would be to search for its spin-triplet 
partner $\psi_Q$ which is almost degenerate with $\eta_Q$. This state is analogous 
to $J/\psi$ in the charmonia and has $J^{PC} = 1^{--}$. 
Here, we discuss the decay and production of a color-singlet spin-triplet $\psi_Q$.
Due to its quantum numbers, $\psi_Q$ does not decay into two gluons and
two h-gluons. It can decay into $ggg$, $g_h g_h g_h$, $gg\gamma$,
$g_h g_h \gamma$, or a pair of fermions via a virtual photon or $Z$ boson.
Because of the singlet nature of $Q$ and $J^{PC} = 1^{--}$, $\psi_Q$ does not decay into
two EW gauge bosons if the $SU(2)_L \times U(1)_Y$ gauge symmetry remains unbroken. 
We find that $\psi_Q$ can decay into $WW$ due to small effects of EW symmetry breaking,  
but the branching ratio of $\psi_Q\to WW$ is quite small.

The decay rates of the $\psi_Q$ into the $ggg$ and $l^+ l^-$ ($l=e,\mu,\tau$) 
final states are given by
\begin{eqnarray}
\Gamma ( \psi_Q\to g g g) &=&
\frac{(\pi^2-9) \alpha_s^3}{36 \pi m_Q^2}
\frac{N_h (N_c^2-1)(N_c^2-4)}{N_c^2}
\left|R_{1S}(0)\right|^2,
\label{psiQggg}%
\\
\Gamma ( \psi_Q\to l^+ l^- ) &=&
\frac{N_c N_h \alpha^2 e_Q^2}{3 m_Q^2}
\left[
1-\frac{2(1-4 x_w)}{(4-r_Z)(1-x_w)}
+\frac{2(1-4 x_w + 8 x_w^2)}{(4-r_Z)^2 (1-x_w)^2}
\right]
\left|R_{1S}(0)\right|^2 .
\end{eqnarray}
The decay rate for $\psi_Q \to g_h g_h g_h$ is given by Eq.~(\ref{psiQggg})
by replacing $\alpha_s$, $N_h$, and $N_c$ by $\alpha_h$, $N_c$, and $N_h$, respectively.
We consider cases in which this decay channel is allowed or kinematically closed.
Note that $\psi_Q\to g_h g_h \gamma$ is also possible if the mass of
the scalar h-glueball is less than $M_{\psi_Q}$.
The decay rates for other channels will be presented in Ref.~\cite{progress}.  
The branching ratios for $\psi_Q$ strongly depend on $\alpha_h$,
and $\psi_Q\to g_h g_h g_h$ or $ g_h g_h \gamma$ becomes a dominant decay
channel for $\alpha_h \gtrsim 0.2\textrm{-}0.3$. However, for $\alpha_h \sim 0.1$,
$\psi_Q \to l^+ l^-$ is dominant, and its branching ratio is about $0.3$~\cite{progress}.
Therefore, the dilepton production via the $\psi_Q$ resonance
would be another promising channel for probing or constraining this model
for smaller $\alpha_h$.
We also note that the search for a new resonance in dijet events 
can constrain this model via $pp\to \psi_Q \to q\bar{q}$.

As is well known, the $\psi_Q$ resonance is strongly constrained by the 
Drell-Yan (DY) production of $q\bar{q} \to \psi_Q \to l^+ l^-$ in $pp$ collisions
with the following cross section:
\begin{eqnarray}
\sigma_{\rm DY}(pp \to \psi_Q \to l^+ l^-) &  = & 
\frac{(2J_{\psi_Q}+1)\Gamma(\psi_Q\to l^+ l^-)}{s \, M_{\psi_Q} \Gamma_{\psi_Q}}
 \times  \sum_{q\bar{q}}^{} C_{q\bar{q}} \Gamma(\psi_Q\to q\bar{q}),
\label{qqxsec}%
\end{eqnarray}
where $C_{q\bar{q}}$ is given by~\cite{Franceschini:2015kwy}
\begin{equation}
C_{q\bar{q}}=\frac{4\pi^2}{9}\int_{M^2/s}^{1} \frac{d \tau}{\tau} 
\left[f_q(\tau)f_{\bar{q}}(M^2/s\tau)
+f_{\bar{q}}(\tau)f_{q}(M^2/s\tau)\right].
\end{equation}
Here, $f_{q,\bar{q}}$ denote the parton distribution functions of $q$
and $\bar{q}$ evaluated at the scale $\mu=M_{\psi_Q}$, and 
$J_{\psi_Q}=1$ is the spin of $\psi_Q$.
For example, by making use of the MSTW2008NLO data~\cite{mstw},
at $\sqrt{s}=13$~TeV, one obtains $C_{u\bar{u}}=1054$, $C_{d\bar{d}}=627$,
$C_{s\bar{s}}=83$, $C_{c\bar{c}}=36$,
and $C_{b\bar{b}}=15.3$ for $\mu=750$~GeV;
and $C_{u\bar{u}}=14.9$, $C_{d\bar{d}}=7.1$,
$C_{s\bar{s}}=0.33$, $C_{c\bar{c}}=0.11$,
and $C_{b\bar{b}}=0.044$ for $\mu=2$~TeV.
In dijet production, $\Gamma(\psi_Q\to l^+ l^-)$ is replaced by
$\sum\Gamma(\psi_Q\to q\bar{q})$ in Eq.~(\ref{qqxsec}).

\begin{figure}[t]
\includegraphics[width=0.45\textwidth]{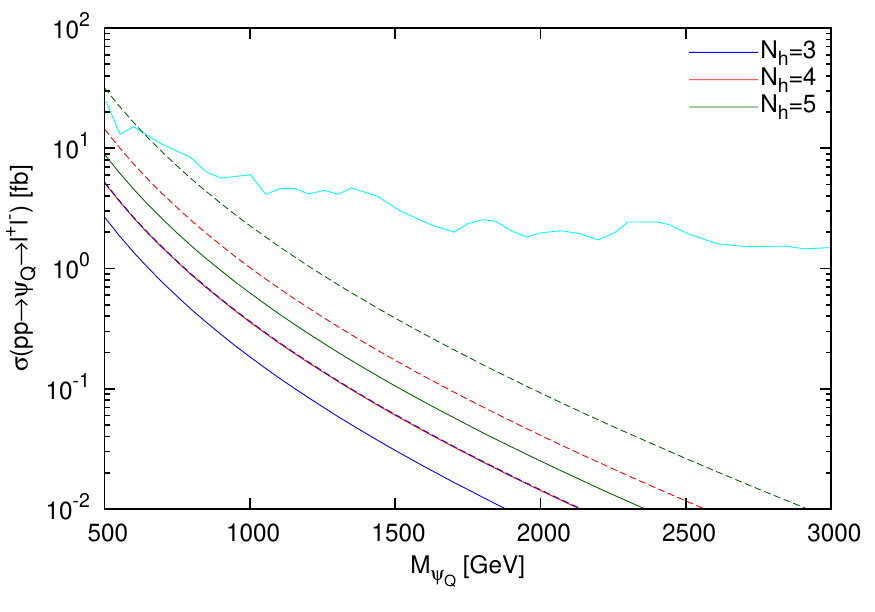}
\includegraphics[width=0.45\textwidth]{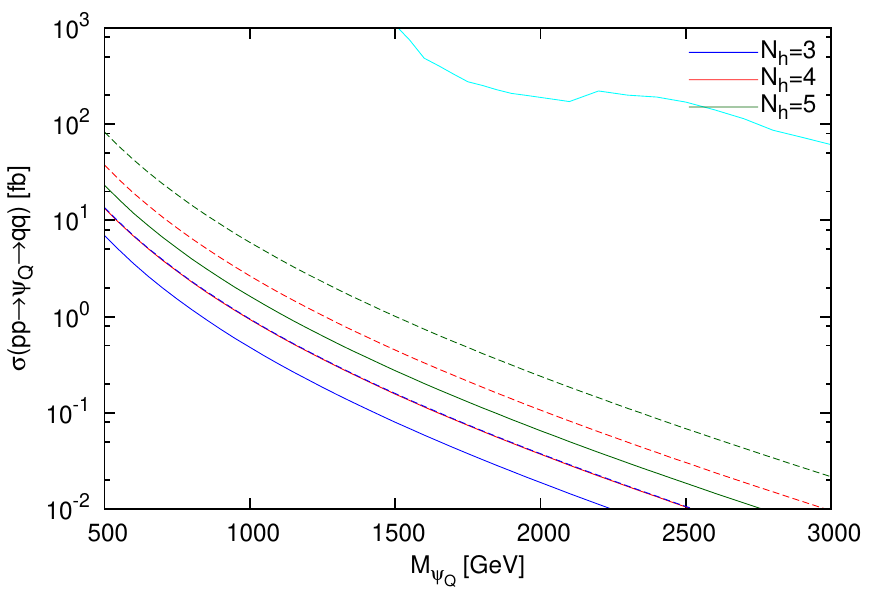}
\caption{
The cross sections for (a) $pp \to \psi_Q \to l^+ l^-$
and (b) $pp\to \psi_Q \to q\bar{q}$
in units of fb for $\alpha_h=0.2$ as functions of $M_{\psi_Q}$ 
at the LHC with $\sqrt{s}=13$~TeV.
The solid (dashed) lines correspond to the case in which
$\psi_Q\to g_h g_h g_h$ is allowed (forbidden).
See the text for explanation of the cyan lines.
}
\label{drellyan}
\end{figure}

In Fig.~\ref{drellyan}(a), the cross section for the DY process,
$pp\to \psi_Q \to l^+ l^-$ ($l=$ either $e$ or $\mu$),
for $\alpha_h=0.2$ at $\sqrt{s} = 13$ TeV  is shown in solid (dashed) lines
in the case in which $\psi_Q\to g_h g_h g_h$ is allowed (forbidden).
The cyan line denotes the upper 95\% C.L. limit on the cross section
times the branching ratio to two leptons at $\sqrt{s}=13$~TeV 
in ATLAS data~\cite{atlasll}.
As shown in Fig.~\ref{drellyan}(a), the $\psi_Q$ production is not constrained
by the DY process except in the region in which $M_{\psi_Q}\lesssim 700$~GeV
and $N_h=5$ when $\psi_Q\to g_h g_h g_h$ is forbidden.

In Fig.~\ref{drellyan}(b), we show the dijet production cross section
in $pp\to \psi_Q\to q\bar{q}$ at $\sqrt{s}=13$~TeV.
The cyan line corresponds to the same upper bound as in Fig.~\ref{figmq}(a) 
with the lepton pair's branching ratio replaced by the light quark pair's branching ratio.
The search for a new resonance in the dijet production does not constrain
this model yet.

\subsection{Excited states}

Another characteristic feature of any composite model is the existence of excited states,
similar to $\psi^{\prime}$, $\eta_c^{'}$, $\Upsilon (nS)$, and so on. These excited states can cascade-decay into  the ground state(s) by emitting h-gluons, gluons, and electroweak gauge bosons,
in analogy with $\psi^{\prime} \rightarrow J/\psi \pi \pi$, $\eta_c \gamma$, etc.  All these channels
require detailed information on the bound-state spectra and the wave functions, 
and we will not consider them any further in this paper.

In passing, we briefly mention the decays and the productions of 
an excited state $\eta_Q^\prime$, which is the $2^1S_0$ state.
We find that the cross section for $pp\to \eta^{\prime}_Q \to \gamma\gamma$
could be about 12\% of that for $pp\to \eta_Q\to \gamma\gamma$.

\begin{figure}[t]
\includegraphics[width=0.45\textwidth]{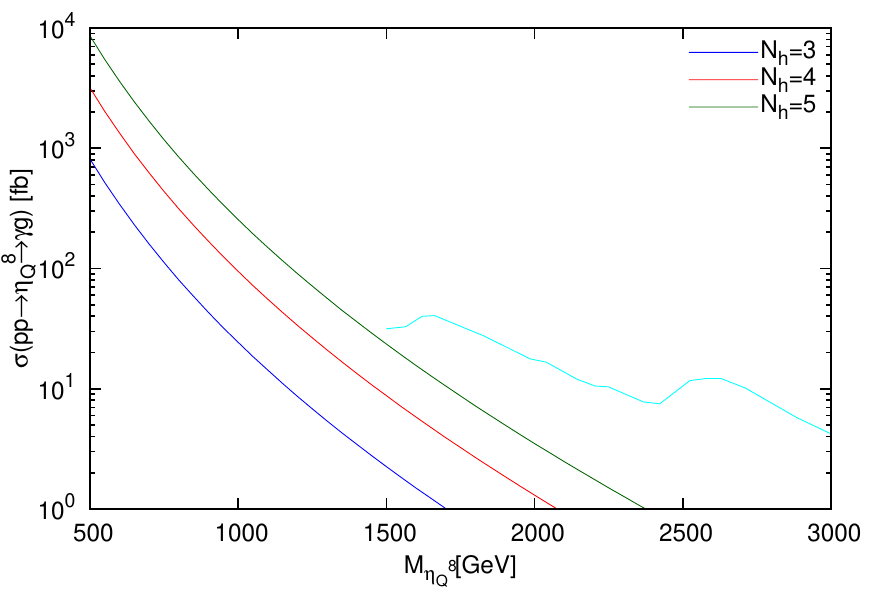}
\includegraphics[width=0.45\textwidth]{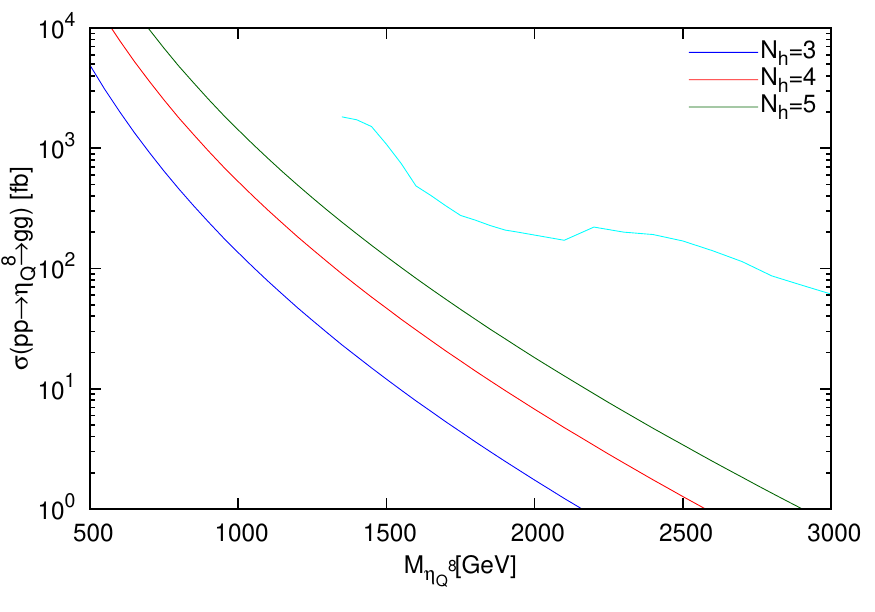}
\caption{
The cross sections for (a) $pp \to \eta_Q^8 \to \gamma g$
and (b) $pp\to \eta_Q^8 \to gg$
in units of fb for $\alpha_h=0.2$ as functions of $M_{\eta_Q^8}$ 
at the LHC with $\sqrt{s}=13$~TeV. See the text for explanation
of the cyan lines.
}
\label{figoctet}
\end{figure}

\subsection{Production and decay of the color-octet bound state}

In this section, we consider the production and decay of the color-octet
bound state, $\eta_Q^8$, which could be formed when the h-color-singlet interaction 
of $Q\bar{Q}$ is much stronger than the color-octet QCD interaction.

$\eta_Q^8$ can decay into two-body modes $gg$, $g\gamma$, $Z g$ and 
three-body modes $ggg$, $gg\gamma$,  as well as $g g_h g_h$ (if kinematically allowed). 
Note that it does not decay into $\gamma\gamma$ or $g_h g_h$ due to color conservation. 
Also, $\eta_Q^8 \rightarrow g \gamma$ is the unique signature for the color-octet bound state, 
unlike  the usual color-singlet bound states. The final state $\gamma +$jet is the same as the final state of 
the excited quark decay $q^* \rightarrow q \gamma$, so the bounds from the excited quark searches 
would apply here.
The three-body modes are suppressed by phase space and will be treated elsewhere~\cite{progress}. The decay rates of $\eta_Q^8 \to gg$, $\gamma g$, and $Zg$ are 
\begin{eqnarray}
\Gamma[\eta_Q^8\to gg] &=&
\frac{(N_c^2-1)(N_c^2-4)N_h \alpha_s^2}{64 N_c m_Q^2} 
\left| R_{\eta_Q}^8 (0) \right|^2, \\
\Gamma[\eta_Q^8\to \gamma g] &=&
\frac{(N_c^2-1)N_h \alpha_s\alpha e_Q^2}{8 m_Q^2} 
\left| R_{\eta_Q}^8 (0) \right|^2, \\
\Gamma[\eta_Q^8\to Zg] &=&
\frac{(N_c^2-1)N_h \alpha_s\alpha e_Q^2 x_w (4-r_Z)}{32(1-x_w) m_Q^2} 
\left| R_{\eta_Q}^8 (0) \right|^2.
\end{eqnarray}
The branching ratios in each of the above decay channels are $0.70$, $0.15$, and $0.15$, respectively.

The production of the color-octet bound state can be constrained by
resonance searches in the dijet production 
corresponding to the $pp\to\eta_Q^8\to gg$ mode,
and in the $\gamma$+jet production 
corresponding to the $pp\to \eta_Q^8\to\gamma g$ mode.
In Fig.~\ref{figoctet}, we depict the cross sections
for Fig.~\ref{figoctet}(a) $pp\to \eta_Q^8\to \gamma g$ and
Fig.~\ref{figoctet}(b) $pp\to \eta_Q^8\to g g$ by setting $\alpha_h=0.2$
as functions of $M_{\eta_Q^8}$.
The cyan line in Fig.~\ref{figoctet}(a) denotes the 95\% C.L. limit 
on the production cross section
times the branching ratio to a photon and a quark or a gluon 
for an excited quark $q^\ast$ at $\sqrt{s}=13$~TeV in ATLAS data~\cite{atlaspg}.
Similar limits can be obtained from the bounds
for the excited quark production at $\sqrt{s}=8$ or $13$~TeV
in CMS or ATLAS data~\cite{gp1,gp2,gp3,gp4}.
The cyan line in Fig.~\ref{figoctet}(b) is the same as that in Fig.~\ref{figmq}(b).
As shown clearly in Fig.~\ref{figoctet}, both production modes at $\sqrt{s}=13$~TeV
do not constrain this model for $\alpha_h=0.2$.
However, for larger $\alpha_h$, this model would be constrained,
in particular, in the $\gamma g$ production channel.

\section{Bound states of scalar hyperquarks}
\label{sec4}

In this section, we consider extra scalar quark singlet $\widetilde{Q} $ with $Y=e_Q=2/3$
and mass $m_{\widetilde Q}$.     
$\widetilde Q$ belongs to the fundamental representation of $SU(N_h)$ gauge theory like $Q$.
The lowest bound state is denoted as 
$\eta_{\widetilde Q}$, which is a color as well as a hypercolor singlet 
bound state of $\widetilde{Q}\widetilde{Q}^\dagger$  in the $S$-wave 
state $\eta_{\widetilde Q}(^1S_0)$  with $J^{PC} = 0^{++}$.  
There will be no analogy of $\psi_Q$ ($^3S_1$) if the constituent particles are scalar quarks
rather than Dirac fermions. Instead, the $J^{PC} =1^{--}$ state ($\chi_{\widetilde Q}$) 
arises from higher radial excitation with nonzero 
orbital angular momentum, $J=L=1$.  
Since the vector resonance for scalar constituents has a zero node at the origin in the radial wave function, the wave function vanishes there.
Its production rate will be suppressed by the derivative of the wave function, 
and thus it will be relatively smaller than the $S$-wave ground state.

\subsection{Productions and decays of $\eta_{\widetilde Q}$, $\chi_{\widetilde Q}$, and $\eta^8_{\widetilde Q}$} 

The scalar bound state $\eta_{\widetilde Q}$ of new scalar h-quarks can decay into two photons, 
$\gamma Z$, $ZZ$, two gluons, or two h-gluons.
The decay widths of these modes are given by
\begin{eqnarray}
\Gamma (\eta_{\widetilde Q} \to \gamma\gamma)&=&
\frac{N_c N_h \alpha^2 e_Q^4}{2 m_Q^2}
\left|\widetilde{R}_{1S}(0)\right|^2,
\\
\Gamma (\eta_{\widetilde Q} \to \gamma Z)&=& 
\frac{N_c N_h \alpha^2 e_Q^4 x_w (4-r_Z)}{4 m_Q^2 (1-x_w)^2}
\left|\widetilde{R}_{1S}(0)\right|^2,
\\
\Gamma (\eta_{\widetilde Q} \to Z Z)&=& 
\frac{N_c N_h \alpha^2 e_Q^4 x_w^2 (8-8 r_Z+3 r_Z^2) \sqrt{1-r_Z}}
{4 m_Q^2 (2-r_Z)^2 (1-x_w)^2}
\left|\widetilde{R}_{1S}(0)\right|^2,
\\
\Gamma (\eta_{\widetilde Q} \to gg) &=&
\frac{N_h (N_c^2-1) \alpha_s^2}{8 N_c m_Q^2} 
\left|\widetilde{R}_{1S}(0)\right|^2,
\\
\Gamma (\eta_{\widetilde Q} \to g_hg_h) &=&
\frac{N_c (N_h^2-1) \alpha_h^2}{8 N_h m_Q^2} 
\left|\widetilde{R}_{1S}(0)\right|^2,
\label{etasQtogg}
\end{eqnarray}
where $\widetilde{R}_{1S}(0)$ is the wave function at the origin of the scalar
quark bound state. Note that $\widetilde{R}_{1S}(0)$ is the same as 
$R_{1S}(0)$ up to one-loop-order correction 
for the QCD-like potential~\cite{Moxhay:1985dy} and the hyperfine splitting, which is absent 
in the case of the scalar h-quark. 
We note that $\eta_{\widetilde Q}$ does not decay into
a pair of fermions or $WW$, just like the case of $\eta_Q$. 

The branching ratios strongly depend on $\alpha_h$ 
if $\eta_{\widetilde Q}\to g_h g_h$ is allowed.
For $\alpha_h\sim \alpha_s$ and $N_h=3$, 
both $BR(\eta_{\widetilde Q}\to g_h g_h)$ and $BR(\eta_{\widetilde Q}\to gg)$
approach $0.5$. However, for $\alpha_h \gtrsim 0.2$,
$BR(\eta_{\widetilde Q}\to g_h g_h)$ becomes dominant over other decay channels.
Actually, $BR(\eta_{\widetilde Q}\to g_h g_h) \gtrsim 0.8$ for $\alpha_h=0.2$~\cite{progress}.
On the other hand, if $\eta_{\widetilde Q}\to g_h g_h$ is kinematically closed,
$BR(\eta_{\widetilde Q}\to gg)$ becomes more than $0.98$ in the entire
parameter space.

\begin{figure}[t]
\includegraphics[width=0.45\textwidth]{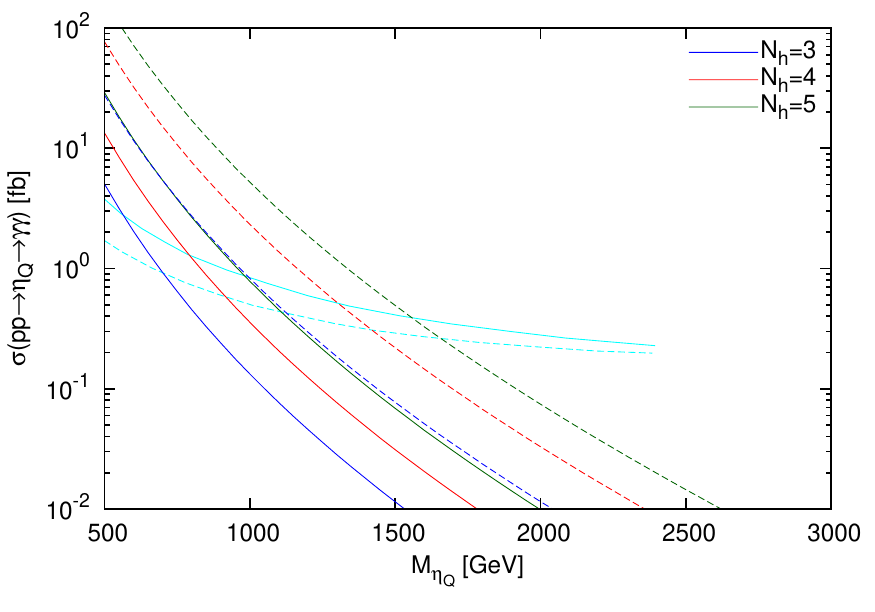}
\includegraphics[width=0.45\textwidth]{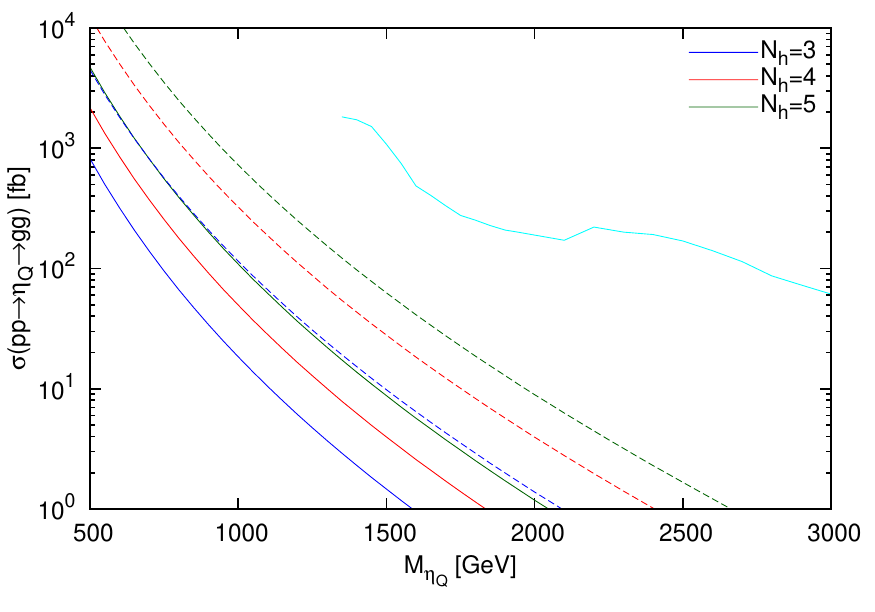}
\includegraphics[width=0.45\textwidth]{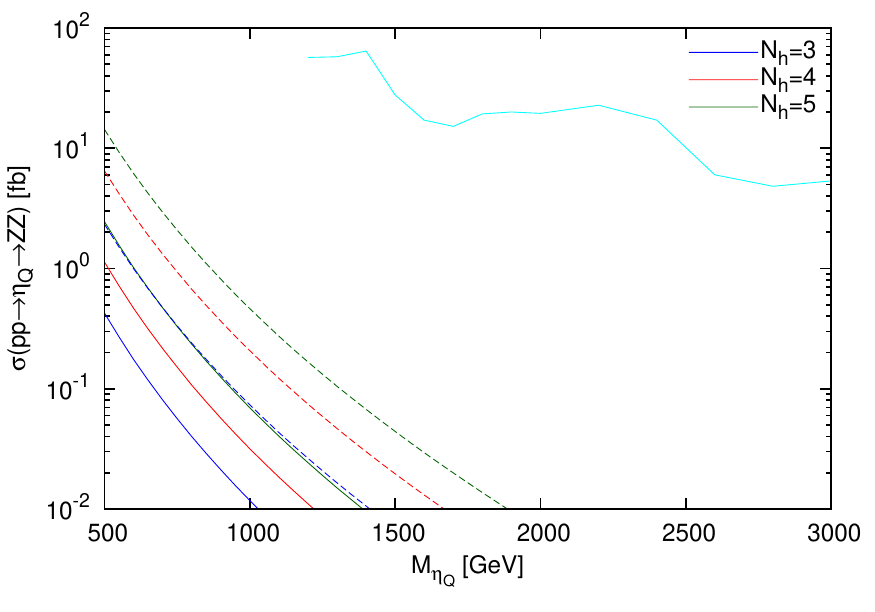}
\includegraphics[width=0.45\textwidth]{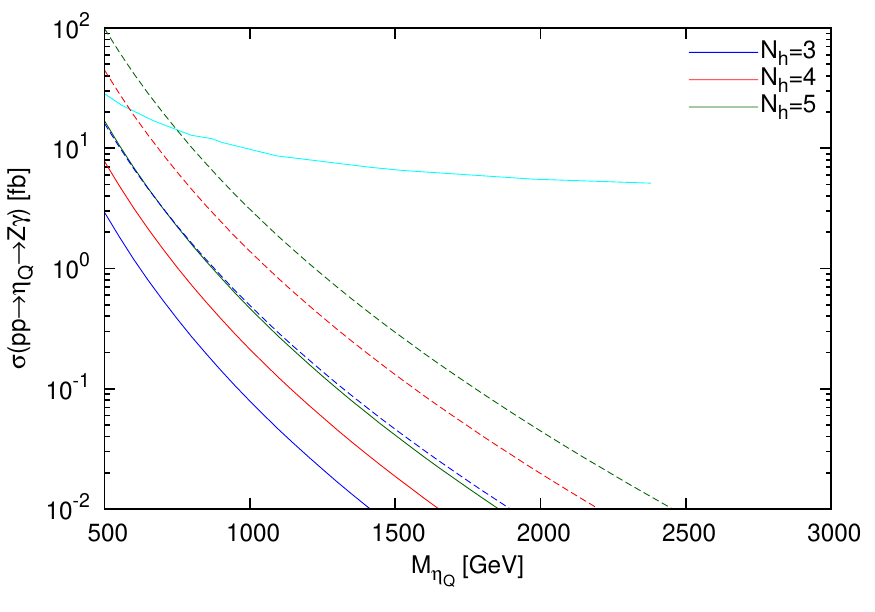}
\caption{
The cross sections for (a) $pp\to \eta_{\widetilde Q} \to \gamma\gamma$,
(b) $pp\to \eta_{\widetilde Q} \to gg$, (c) $pp\to \eta_{\widetilde Q} \to ZZ$,
and (d) $pp\to \eta_{\widetilde Q} \to Z\gamma$ for $\alpha_h=0.2$
at the LHC with $\sqrt{s}=13$~TeV in units of fb 
as functions of $M_{\eta_{\widetilde Q}}$.
The cyan lines are the same experimental upper bounds as in Fig.~\ref{figmq}.
}
\label{figmqs}
\end{figure}

In Figs.~\ref{figmqs}(a)--\ref{figmqs}(d), we plot the cross sections for
$pp\to \eta_{\widetilde Q}\to \gamma\gamma$,
$pp\to \eta_{\widetilde Q}\to gg$,
$pp\to \eta_{\widetilde Q}\to ZZ$,
and $pp\to \eta_{\widetilde Q}\to Z\gamma$, respectively,
for $\alpha_h=0.2$ at $\sqrt{s}=13$~TeV, as functions of $M_{\eta_{\widetilde Q}}$
with the same experimental upper bounds as in Fig.~\ref{figmq}.
The solid (dashed) lines correspond to the cases in which
the $\eta_{\widetilde Q}\to g_h g_h$ decay is allowed (forbidden).
The cross sections for the $\eta_{\widetilde Q}$ production in Fig.~\ref{figmqs}
are a little bit smaller than those for the $\eta_Q$ production 
in Fig.~\ref{figmq}.
The difference mainly originates in the different spins of the particles
constituting the bound states. However, general features are the same as
in Fig.~\ref{figmq}.

The vector resonance $\chi_{\widetilde Q}$ can decay into a pair of leptons,
and thus it is constrained by the DY process like $\psi_Q$ in the fermion case.
We find that the production cross section for $pp\to \chi_{\widetilde Q}\to l^+l^-$
is highly suppressed by the derivative of the wave function
at the origin.
For $M_{\eta_{\widetilde Q}}>500$~GeV, we find that
$\sigma(pp\to \chi_{\widetilde Q}\to l^+l^-)\lesssim O(10^{-4})$~fb,
which is much smaller than the LHC upper bound.
Similarly, the cross section for the dijet production is
$\sigma(pp\to \chi_{\widetilde Q}\to q\bar{q})\lesssim O(10^{-2})$~fb,
which is not constrained by the data at all.

\begin{figure}[t]
\includegraphics[width=0.45\textwidth]{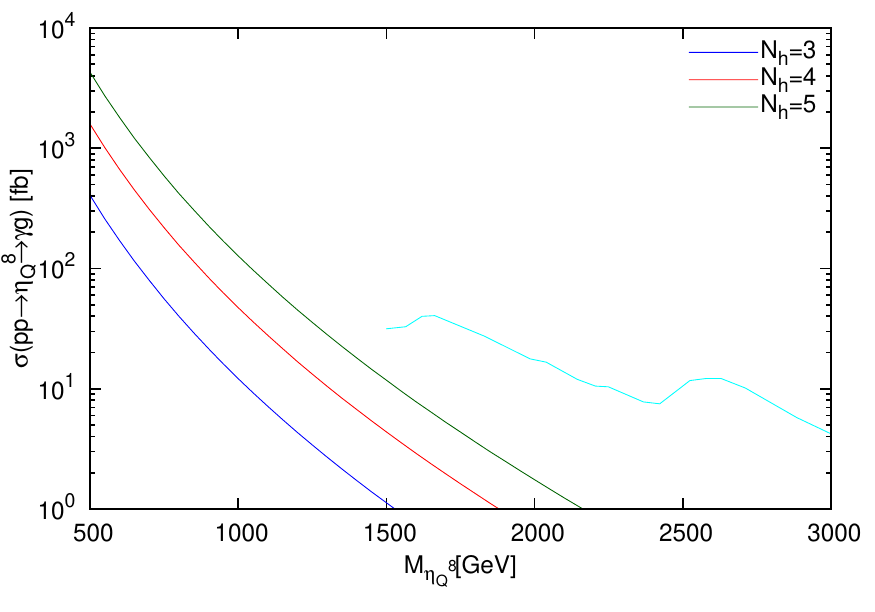}
\includegraphics[width=0.45\textwidth]{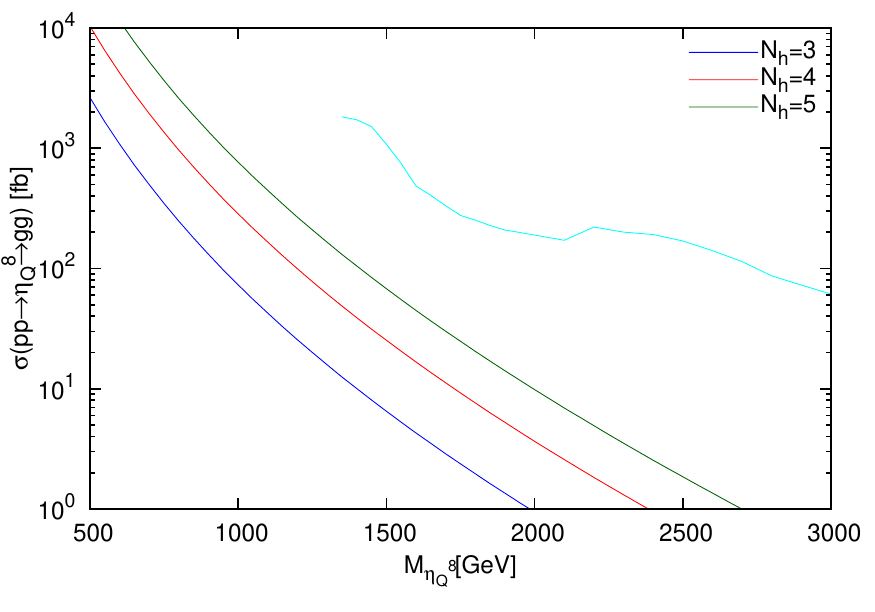}
\caption{
The cross sections for (a) $pp \to \eta_{\widetilde Q}^8 \to \gamma g$
and (b) $pp\to \eta_{\widetilde Q}^8 \to gg$
in units of fb for $\alpha_h=0.2$ as functions of $M_{\eta_{\widetilde Q}^8}$ 
at the LHC with $\sqrt{s}=13$~TeV. The cyan lines are the same experimental upper bounds used in Fig.~\ref{figoctet}.
}
\label{figoctets}
\end{figure}

The scalar h-quarks can also make a QCD color-octet bound state but an h-color singlet.
We denote such a ground state by $\eta_{\widetilde Q}^8$, just
like $\eta_Q^8$ in the h-quark model. 
$\eta_{\widetilde Q}^8$ can decay into $gg$, $g\gamma$, or $Z g$, where
we suppress the three-body decay modes. The decay rates of the two-body modes are given by
\begin{eqnarray}
\Gamma[\eta_{\widetilde Q}^8\to gg] &=&
\frac{(N_c^2-4)N_h \alpha_s^2}{16 N_c m_Q^2} 
\left| R_{\eta_{\widetilde Q}}^8 (0) \right|^2, \\
\Gamma[\eta_{\widetilde Q}^8\to \gamma g] &=&
\frac{N_h \alpha_s\alpha e_Q^2}{2 m_Q^2} 
\left| R_{\eta_{\widetilde Q}}^8 (0) \right|^2, \\
\Gamma[\eta_{\widetilde Q}^8\to Zg] &=&
\frac{N_h \alpha_s\alpha e_Q^2 x_w (4-r_Z)}{8(1-x_w) m_Q^2} 
\left| R_{\eta_{\widetilde Q}}^8 (0) \right|^2.
\end{eqnarray}
The branching ratios of the above decay channels 
are $0.70$, $0.15$, and $0.15$, respectively.
In Figs.~\ref{figoctets}(a) and \ref{figoctets}(b), we show the production cross sections 
for $pp \to \eta_{\widetilde Q}^8 \to \gamma g$
and $pp\to \eta_{\widetilde Q}^8 \to gg$, respectively,
in units of fb for $\alpha_h=0.2$ as functions of $M_{\eta_{\widetilde Q}^8}$ 
at $\sqrt{s}=13$~TeV, 
compared with the same experimental bound (cyan lines) used in Fig.~\ref{figoctet}.
We find that the expected cross sections in the scalar h-quark model 
are half of those in the h-quark model, and 
neither channel is constrained by the LHC data at $\sqrt{s}=13$~TeV yet.

\section{How to distinguish a composite $Q \overline Q$ from
${\widetilde Q}{\widetilde Q}^\dagger$?}
\label{sec5}

One of the key questions is how to distinguish  $\eta_Q$ from 
$\eta_{\widetilde Q}$ if one finds a heavy diphoton 
resonance state in the near future at the LHC.
This can be answered by noting that the $J^{PC}$ quantum
numbers of two states are different, namely $0^{-+}$ vs $0^{++}$.  Hence, the polarizations
of two photons in the final states should be orthogonal vs parallel. 
A similar issue has been studied for the 125~GeV Higgs to determine its $J^{PC}$ quantum numbers.  For example, one can study the azimuthal  angle distribution of 
the forward dijet in  $gg \rightarrow \eta_{Q}~({\rm or}~\eta_{\widetilde Q}) \rightarrow \gamma\gamma$.
Furthermore, if the $gg\to \eta_Q~(\textrm{or~} \eta_{\widetilde Q}) \to ZZ$ 
channel is kinematically allowed, one may study the $J^{PC}$ quantum
numbers of the scalar or pseudoscalar resonance via the angular distribution of decay products
of the two $Z$ bosons.

Another possible way to distinguish 
the two composite scenarios is via the DY production of the vector resonance $\psi_Q$ or $\chi_{\widetilde Q} \to l^+l^-$. 
As shown in Fig.~\ref{drellyan}, the predicted cross section for the DY
production of $\psi_Q\to l^+l^-$ is $0.1\sim 1$~fb at $\sqrt{s}=13$~TeV.
On the other hand, we find that the cross section for the DY production of
$\chi_{\widetilde Q}\to l^+l^-$ is at most $10^{-4}$~fb at $\sqrt{s}=13$ TeV.
Therefore, the two ratios 
\begin{equation}
\frac{\sigma(pp \rightarrow \psi_Q \rightarrow l^+ l^-)}{\sigma(pp\rightarrow \eta_Q \rightarrow \gamma\gamma)} 
~~\textrm{vs}~~ 
\frac{\sigma(pp \rightarrow \chi_{\tilde{Q}} \rightarrow l^+ l^-)}{\sigma(pp\rightarrow \eta_{\tilde{Q}} \rightarrow \gamma\gamma)} \; ,
\end{equation}
in which some unknown factors such as $N_h$ and the wave functions at the origin are canceled out,  may prove to be useful in distinguishing between the two cases.

\section{Interpretation of diphoton and photon + jet resonances 
as composite scalar or pseudoscalar at the LHC}
\label{sec6}

Although there is no significant clue on any new physics at the LHC,
there are a few resonant excesses with small significances deviated from SM predictions.
In this section, we investigate the possibility that
these small excesses might be interpreted as pseudoscalar or scalar composite particles,
whose constituents are either new vectorlike quarks ($Q\overline{Q}$) or scalar quarks 
($\widetilde{Q} \widetilde{Q}^\dagger$).
In this section, we fix $N_h=3$, but we set $\alpha_h$ and $e_Q$ to be free.

\subsection{Two diphoton resonances at 710~GeV and 1.6~TeV}

AT the 2016 ICHEP conference, both ATLAS~\cite{ATLAS:2016eeo} and CMS~\cite{Khachatryan:2016yec} reported
new results on the 750~GeV diphoton excess, including new data in 2016. 
Combining the 2015 and 2016 data, the ATLAS Collaboration~\cite{ATLAS:2016eeo} observed a local significance of $2.3\sigma$ excess at 
$710$~GeV with a large decay width to mass ratio, $\Gamma/M=0.1$, 
and another one of $2.4\sigma$ at $1.6$~TeV with a narrower width.
On the other hand, the CMS Collaboration~\cite{Khachatryan:2016yec} has observed
no significant excess by combining 2015 and 2016 data.
In this section, we attempt to identify these small excesses in ATLAS data 
as signals of a pseudoscalar or scalar composite particle in the hypercolor model.

\begin{figure}[t!]
\includegraphics[width=0.45\textwidth]{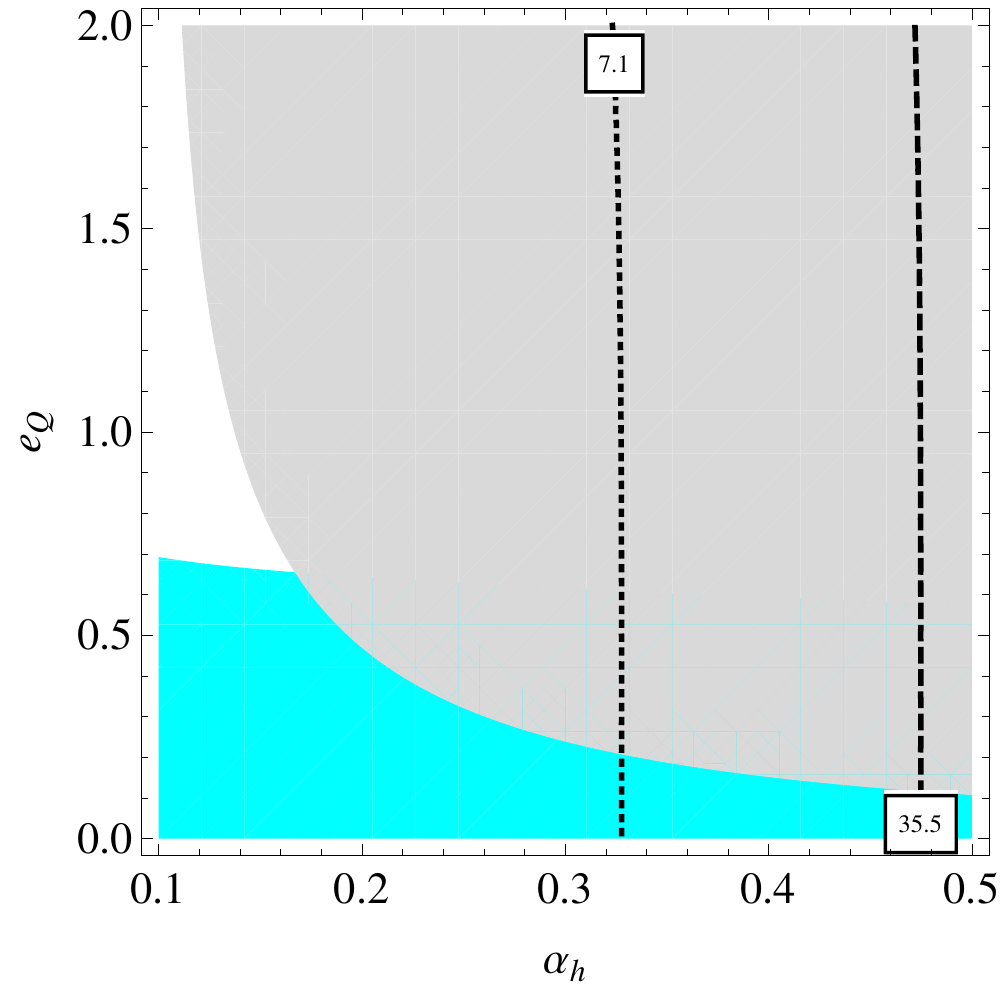}
\includegraphics[width=0.45\textwidth]{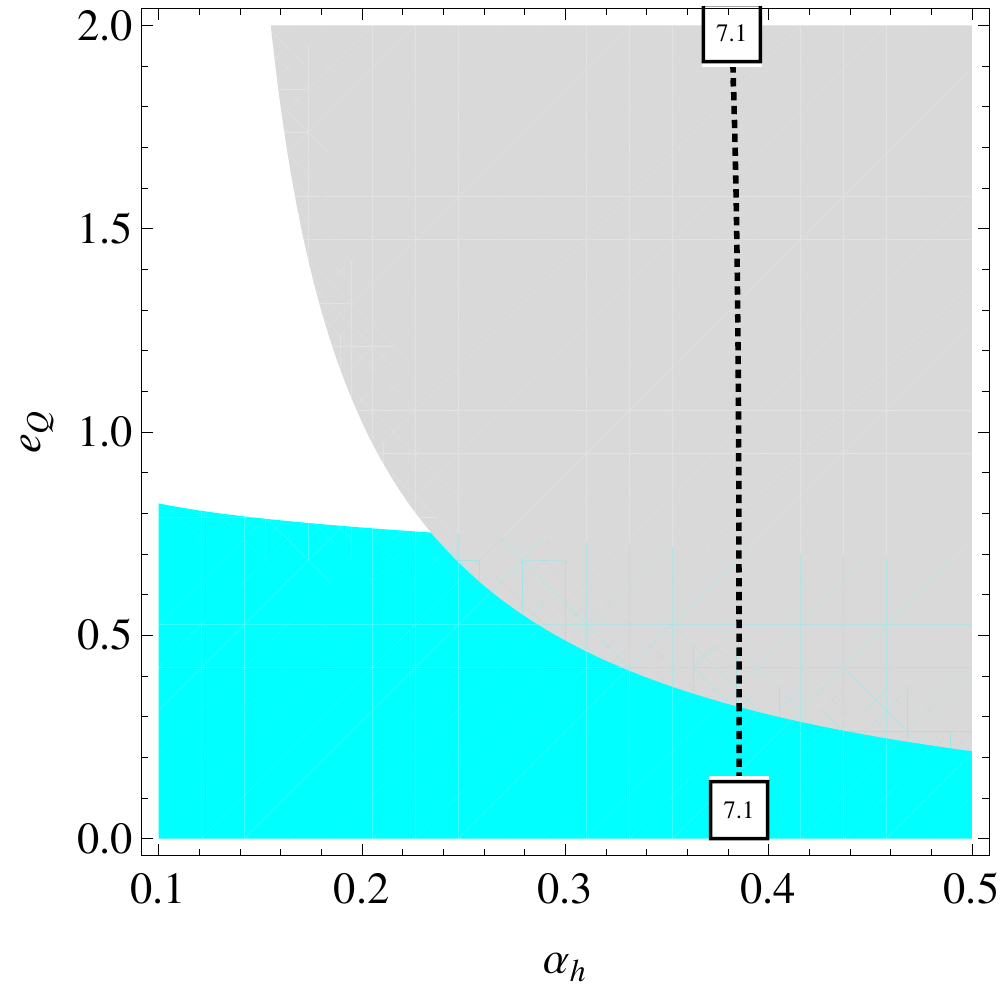}
\caption{
The allowed region of $\alpha_h$ and $e_Q$ for a resonance at $710$~GeV
in the diphoton channel at ATLAS~\cite{ATLAS:2016eeo}. 
The left (right) panel corresponds to the h-(scalar) quark model.
The gray region is ruled out by the photon+jet search
at $\sqrt{s}=8$~TeV in ATLAS data~\cite{gp1}. 
The dashed (dotted) line denotes the total decay width 
$\Gamma_{\eta_{Q/{\widetilde Q}}}$/GeV
corresponding to the ratio $\Gamma/M=0.05$ $(0.01)$.
}
\label{fig710}
\end{figure}

First, we consider the excess at $710$~GeV. According to the MSTW2008NLO data~\cite{mstw}, 
we have $C_{gg}=2807$ and 237 at $\sqrt{s}=13$  and $8$~TeV, respectively.
The expected value for the $\gamma\gamma$ signal 
at $710$~GeV in the SM is about $1$~fb,
and it could reach about $2$~fb with $2\sigma$ uncertainty~\cite{ATLAS:2016eeo}. 
In the following analysis, we interpret the $2.3\sigma$ local excess at $710$ GeV
as the production of $\eta_Q$ or $\eta_{\widetilde Q}$ decaying into 
$\gamma\gamma$, whose signal strength is taken to be less than $1.3$~fb. 

In Fig.~\ref{fig710}, the cyan region corresponds to the region in which
$\sigma(pp\to \eta_Q~(\textrm{or~} \eta_{\widetilde Q})\to \gamma\gamma)<1.3$~fb when $\eta_{Q/\widetilde Q}\to g_h g_h$ is allowed for the $710$~GeV resonance.
The gray region is ruled out by the bound from 
the search for a resonance decaying into a photon + jet at $\sqrt{s}=8$~TeV
in ATLAS data~\cite{gp1}. Explicitly, we set the bound
$\sigma(pp\to \eta_{Q / {\widetilde Q}}^8\to \gamma g) < 18$~fb for $M_{\eta_{Q/\widetilde Q}}\sim 710$~GeV and 
$\Gamma / M\sim 5$\% by assuming that 
the product of the efficiency and acceptance is $0.33$~\cite{Kats:2016kuz}. 
The dashed (dotted) line denotes the total decay width 
$\Gamma_{\eta_{Q/{\widetilde Q}}}/$GeV
corresponding to the ratio $\Gamma/M=0.05$ $(0.01)$.
The left (right) panel in Fig.~\ref{fig710} corresponds to the case of
$\eta_Q$ ($\eta_{\widetilde Q}$).
In the h-scalar quark model, the allowed region is a little bit broader
than in the h-quark model.
Both models prefer the narrow decay width for the resonance so that
the bound from the $\gamma+$ jet search might become stronger.
There would be other constraints from the dijet, dilepton, $ZZ$, and $Z\gamma$
searches, but the constraints are much weaker than for the photon+jet search,
as shown in Fig.~\ref{figmq}.

\begin{figure}[t!]
\includegraphics[width=0.45\textwidth]{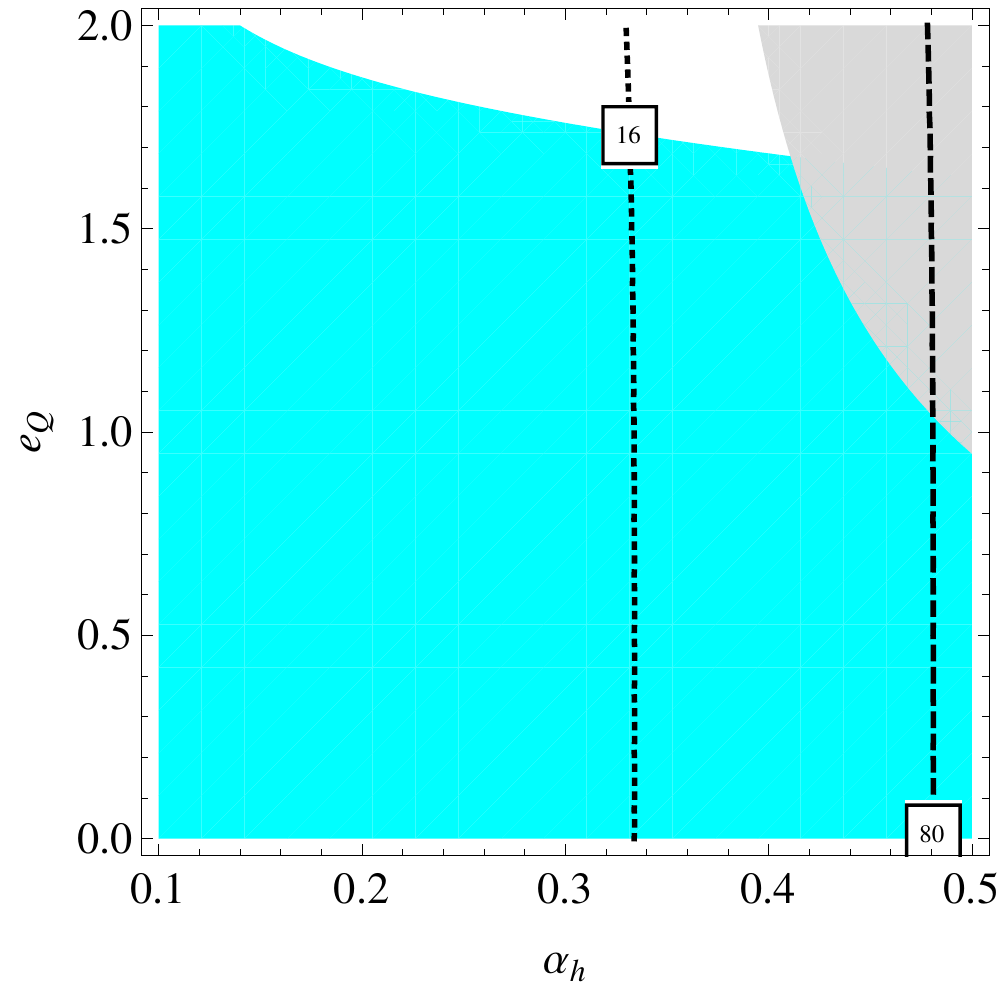}
\includegraphics[width=0.45\textwidth]{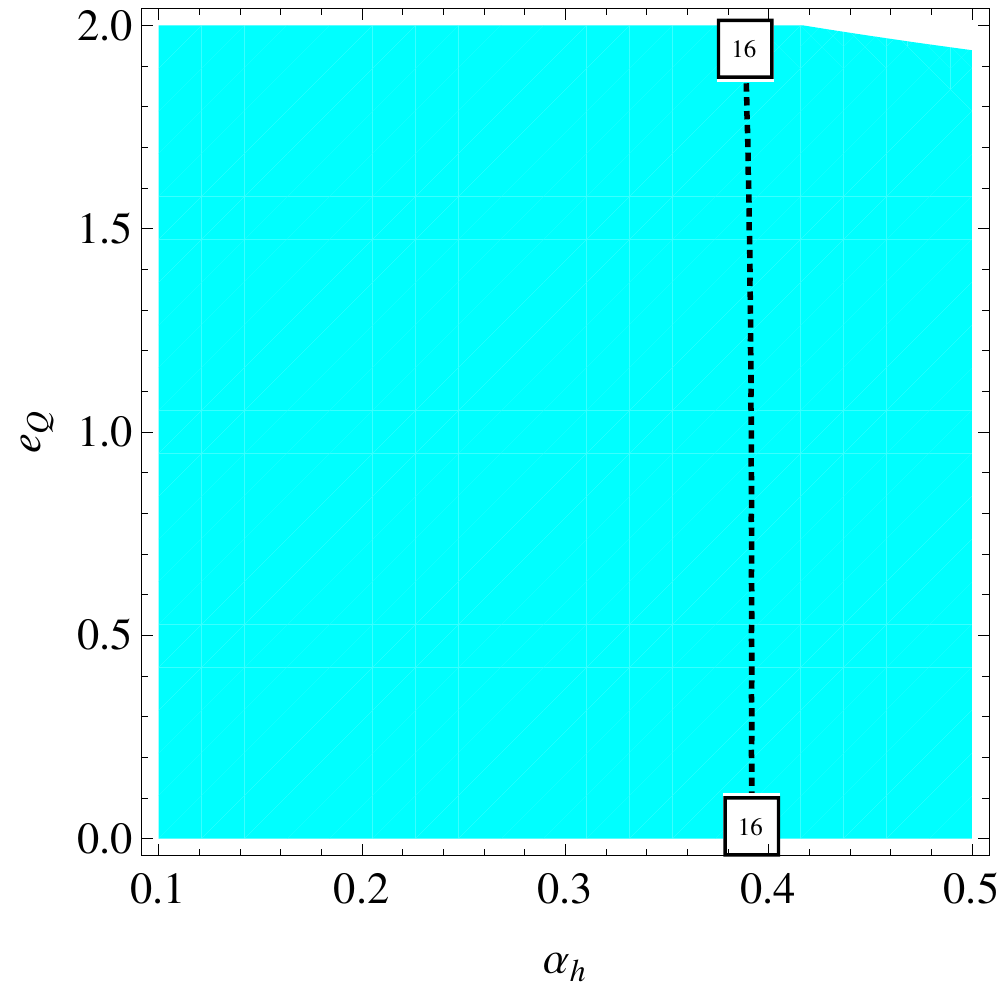}
\caption{
Same as Fig.~\ref{fig710}, for the 1.6~TeV resonance.}
\label{fig16}
\end{figure}

Next, we consider the excess at $1.6$~TeV.  Here, $C_{gg}=31.05$ and 1.18
at $\sqrt{s}=13$ and $8$~TeV, respectively.
The expected value for the cross section times the branching ratio
to $\gamma\gamma$ in the SM at $1.6$~TeV is about $0.3$~fb, 
and it could reach about $0.8$~fb with $2\sigma$ uncertainty~\cite{ATLAS:2016eeo}. 
Therefore, we interpret the $2.4\sigma$ local excess at $1.6$~TeV
as the production of $\eta_Q$ or $\eta_{\widetilde Q}$ decaying into 
$\gamma\gamma$, whose cross section is less than $0.7$~fb. 

In Fig.~\ref{fig16}, the cyan region corresponds to the region in which 
$\sigma(pp\to \eta_Q~(\textrm{or~} \eta_{\widetilde Q})\to \gamma\gamma)<0.7$~fb when $\eta_{Q/\widetilde Q}\to g_h g_h$ is kinematically allowed for the
$1.6$~TeV resonance.
As in Fig.~\ref{fig710}, the gray region is ruled out by the bound from 
the search for a resonance decaying into a photon + jet 
at $\sqrt{s}=8$~TeV in ATLAS data~\cite{gp1}, and we set the bound
$\sigma(pp\to \eta_{Q/\widetilde Q}^8\to \gamma g) < 4.2$~fb for $M_{\eta_{Q/\widetilde Q}}=1.6$~TeV and 
$\Gamma/M= 5$\% by assuming that 
the product of the efficiency and acceptance is $0.33$~\cite{Kats:2016kuz}. 
Compared to the previous resonance at $710$~GeV,
the $1.6$~TeV resonance has a much broader region of the parameter space
and is less constrained by other LHC data.

\begin{figure}[b!]
\includegraphics[width=0.45\textwidth]{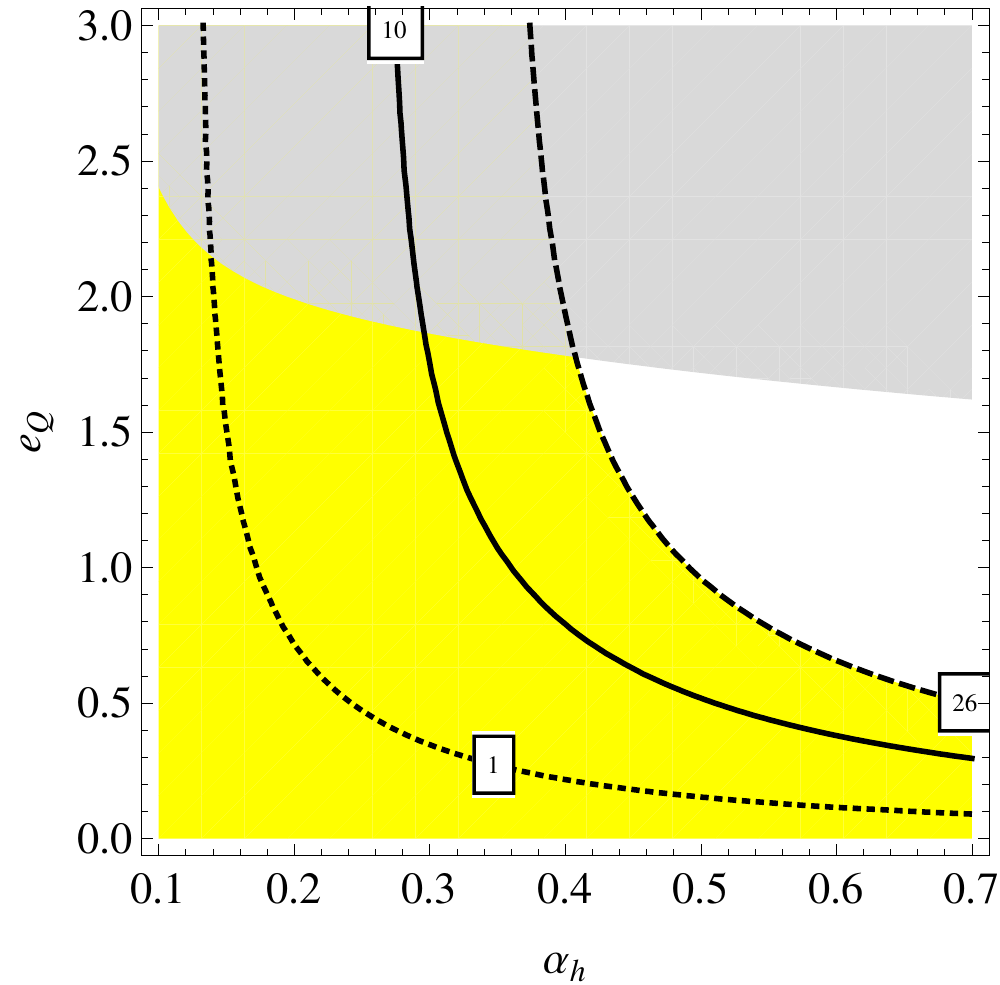}
\includegraphics[width=0.45\textwidth]{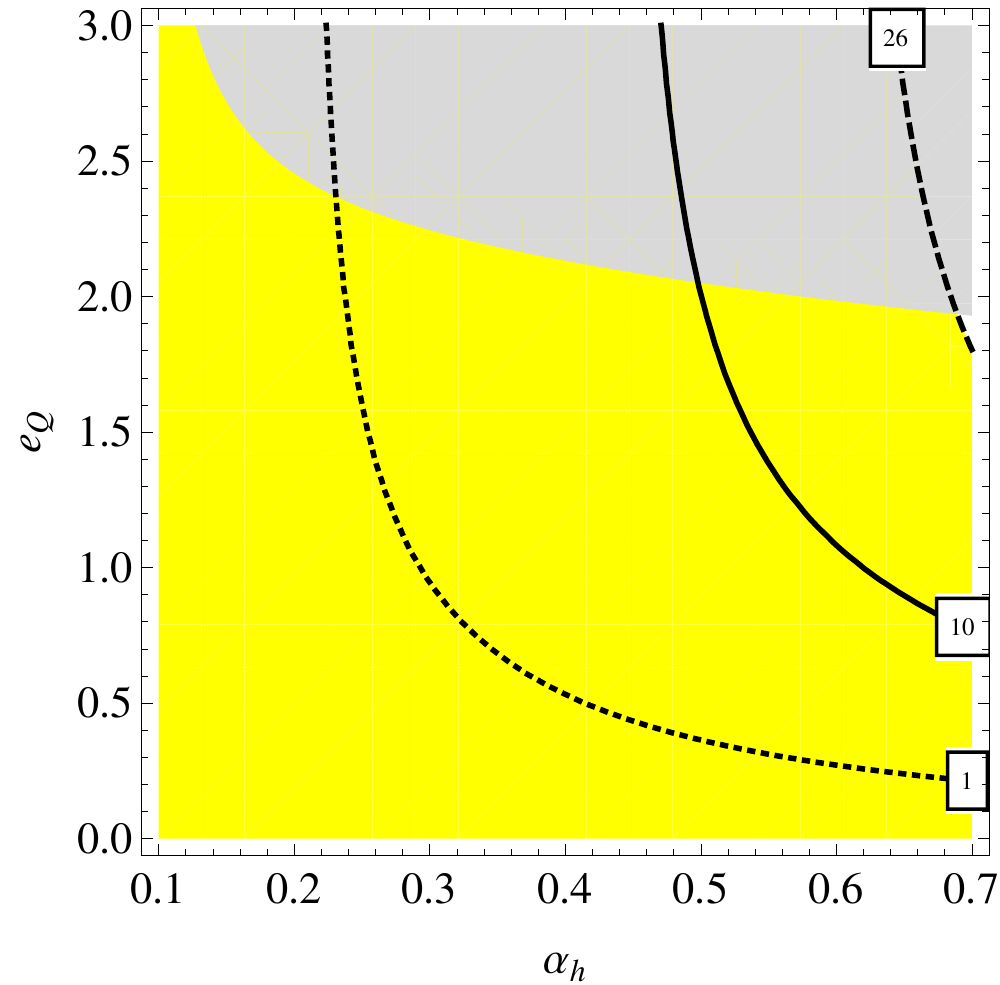}
\caption{
The allowed region of $\alpha_h$ and $e_Q$ for a resonance at $2$~TeV
in the photon+jet search at ATLAS~\cite{ATLAS:2016eeo}. 
The left (right) panel corresponds to the h-(scalar) quark model.
}
\label{fig2000}
\end{figure}

\subsection{$\gamma +$ Jet Resonance at 2 TeV}
 
The CMS Collaboration also announced that there might be some excess
around $2$~TeV in the photon+jet channel~\cite{gp2}.
The largest deviation is seen at a mass of $2.0$~TeV with a 
cross section about $45$~fb, while the SM background expectation 
is about $19$~fb~\cite{gp2}. 
Here, we interpret the excess as the production of the color-octet state, $\eta_{Q/\widetilde Q}^8$ 
decaying into $\gamma g$ for $N_h=3$, whose cross section is restricted to be less than $26$~fb. 

In the left (right) panel of Fig.~\ref{fig2000}, 
the yellow regions denote the allowed regions of $\alpha_h$ and $e_Q$
for the $\eta_Q^8$ ($\eta_{\widetilde Q}^8$) case.
The lines denote the contour values of the cross section for the photon+$g$ production
from the octet states.
The gray regions are disfavored by the diphoton search at $\sqrt{s}=13$~TeV
by assuming $\Gamma/M = 2\%$
in ATLAS data~\cite{ATLAS:2016eeo}, corresponding to 
the region where $\sigma(pp\to \gamma\gamma) > 0.2$~fb.
In the scalar h-quark case (right), a much broader region is allowed, but 
it is impossible to achieve more than $10$~fb for the cross section
in the perturbative region.
However, in the h-quark case (left), it is possible to achieve a cross section of $10$~fb
for $\alpha_h \sim 0.3$ and $e_Q\sim 1.5$.

\section{Conclusion}
\label{sec7}

Diphoton or, in general, diboson resonance  can play the role as a window 
to reveal new physics beyond the SM, like the existence of a hidden strongly interacting sector studied in this work.

In this paper, we have studied the possibility that a high-mass diphoton resonance 
is a composite scalar or pseudoscalar boson made up of $Q\overline{Q}$ or $\widetilde{Q} \widetilde{Q}^\dagger$.
We have calculated the diphoton production cross section $pp \rightarrow \eta_Q (\eta_{\widetilde Q}) 
\rightarrow \gamma \gamma$ and the Drell-Yan production cross section from $pp \rightarrow q\bar{q} 
\rightarrow  \psi_Q (\chi_{\widetilde Q}) \rightarrow l^+ l^-$  at LHC $8$~TeV. 
We found that the Drell-Yan production via $\psi_Q$ at $\sqrt{s} = 8$~TeV 
has already been constrained for the scenario of $Q\overline{Q}$ bound state.   
We discussed how to distinguish the two composite scenarios by determining 
the $J^{PC}$ of the scalar $\eta_{\widetilde Q}$ or pseudoscalar $\eta_Q$ 
diphoton resonance and using the Drell-Yan production of charged leptons 
of the  $\psi_Q$ or $\chi_{\widetilde Q}$ resonance.
The total decay width of $\eta_Q$ or  $\eta_{\widetilde Q}$ can be either large or small depending 
on whether the $g_h g_h$ mode is open or closed. 
We note that the h-glueball case has been omitted in other similar analysis in the literature.

We interpreted the two small diphoton ``excesses'' 
at $710$~GeV and $1.6$~TeV reported by the LHC as the scalar or 
pseudoscalar composite in our model and determined the allowed regions 
of the parameter space from the data.
We also found that existing photon+jet data from ATLAS impose strong constraints on the
color-octet state $\eta^8_{Q}$ or $\eta^8_{\widetilde Q}$.

Besides the hyperquarkonia approach we are adopting here for the diboson resonances, 
there are many alternative composite interpretations as well. For example, in the composite Higgs model~\cite{Belyaev:2016ftv},
the diboson resonances are considered as pNGBs. 
However, there are important distinctions between these two approaches using hyperquarkonia and pNGBs.  
The most notable distinction is that while the hyperquarkonia are formed by new strong confinement force, 
the pNGBs are coming from spontaneous symmetry breaking. Hence the mass differences between the lowest-lying state and excited states are generally quasi-degenerated with mass differences less than $100$~GeV or so in the former case, but large in the latter case. Moreover, in the hyperquarkonia approach, we can consider 
both fermionic and bosonic  constituents in the new gauge group, while only fermionic constituents are 
possible to generate pNGBs. We have showed that one can use Drell-Yan to differentiate these two composite 
scenarios based on fermionic or bosonic constituents.

Finally, we note that for the case of 
h-quarks and scalar h-quarks forming $SU(2)_L$ doublets, 
general diboson resonances and even charged composites as discussed in Ref.~\cite{Cheung:2008ke} are also
possible.
$P$-wave scalar h-quark bound states are also interesting. 
These are all potentially relevant at LHC run II in the searches for new physics.
We hope to report these results in more detail elsewhere~\cite{progress}.

\vfill

\acknowledgments
This work is supported in part by National Research Foundation of Korea (NRF) Research Grant No.~NRF-2015R1A2A1A05001869,  by the NRF grant funded by the Korea government (MSIP) 
(No.~2009-0083526) through the Korea Neutrino Research Center at Seoul National University (P.~K.), and by the Ministry of Science and Technology (MoST) of Taiwan
under Grant No.~101-2112-M-001-005-MY3 (C.~Y. and T.~C.~Y.).
The work of C.~Y. is also supported in part by the Do-Yak project of NRF under Contract
No.~NRF-2015R1A2A1A15054533 and by the Basic Science Research Program through
the National Research Foundation of Korea(NRF) funded by
the Ministry of Science, ICT and Future Planning(No.~2017R1A2B4011946).

\end{document}